\documentclass[epj]{svjour}
\usepackage{graphics}
\begin{document}
\title{Invisible surface defects in a tight-binding lattice}
\author{Stefano Longhi \inst{1} 
}                     
\offprints{}          
\institute{Dipartimento di Fisica, Politecnico di Milano and Istituto di Fotonica e Nanotecnologie del Consiglio Nazionale delle Ricerche, Piazza Leonardo da Vinci 32, I-20133 Milano (Italy)}
\date{Received: date / Revised version: date}
%
\abstract{Surface Tamm states arise in one-dimensional lattices from some defects at the lattice edge and their energy generally falls in a gap of the crystal. The defects at the surface change rather generally the phase of propagative Bloch waves scattered off at the lattice edge, so that an  observer, far from the surface, can detect the existence of edge defects from e.g. time-of-flight measurements as a delay or an advancement of a Bloch wave packet. Here we show that a special class of defects can sustain surface Tamm states which are invisible, in a sense that reflected waves acquire the same phase as in a fully homogeneous lattice with no surface state. Surface states have an energy embedded into the tight-binding lattice band and show a lower than exponential (algebraic) localization. Like most of bound states in the continuum of von Neumann - Wigner type, such states are fragile and decay into resonance surface states in presence of perturbations or lattice disorder. The impact of structural lattice imperfections and disorder on the invisibility of the defects is investigated by numerical simulations.
} 
\maketitle
%

\section{Introduction}

Surface waves localized at an
interface between two different media are ubiquitous in physics  \cite{Davi}. They have been studied in several physical fields ranging form  condensed matter  physics to optics. 
In solid-state physics,  electronic surface waves 
at the edge of a truncated crystal have been commonly classified as either
Tamm \cite{Tamm} or Shockley \cite{S} surface waves, depending on the underlying localization mechanism \cite{Lippmann,Zak}.  Tamm surface states arise from an asymmetrical surface potential, their formation requires exceeding
a threshold perturbation of the surface potential and their energies lie in a gap of the crystal. 
On the other hand, Shockley surface waves appear while the periodic potentials are symmetrically
terminated \cite{S} and result from the crossover of adjacent bands.
Experimental observation of surface states in solids remained elusive
for decades until the advent of semiconductor superlattices \cite{super1,super2}. 
 In optics, surface waves are known since long time, for example they arise at the interface between homogeneous and periodic layered media \cite{Yeh1,Yeh2}.
 Optical analogues of Tamm and Shockley surface
states have been extensively studied and observed for different types of photonic crystals
and waveguide lattices \cite{uff1,uff2,uff3,uff4,uff5,uff6}. 
The study of surface waves in optics benefits from 
the possibility to to explore  the nonlinear domain and thus to observe nonlinear surface states \cite{nl1,nl2,nl3,nl4,nl5,nl6}, which is unfeasible in
electronic systems. Optics also offers the possibility to engineer with great accuracy the surfaces and interfaces and to realize synthetic surfaces that can not be found in nature. In this way, surface Tamm states with lower-than-exponential localization and energies lying in an allowed band of the crystal, i.e. bound states in the continuum (BIC) of von Neumann - Wigner type \cite{Wigner}, have been predicted and experimentally observed in recent works \cite{BIC1,BIC2}.\par
In spite of such a great amount of theoretical and experimental studies on Tamm states, the effects of surface defects on the reflection of Bloch (propagative) states in the crystal have been so far overlooked. If we consider a one-dimensional lattice, the perturbation of the surface potential that is needed to sustain the Tamm (localized) state changes rather generally the phase of the reflected Bloch waves as compared to the unperturbed lattice \cite{cinesi}. This implies that a propagative Bloch wave packet reflected from the surface keeps some information about the existence of the Tamm state, i.e. the surface state does not appear to be 'invisible' from an outsider observer. At first sight, such a property seems to be rather general and can be proven, for example, by considering simple tight-binding lattice models of Tamm states on one-dimensional lattices \cite{book1,book2} (see the following section 2). Indeed, it is known that defects in the {\it bulk} of otherwise periodic and {\it Hermitian} lattices, sustaining localized (defect) modes, are not invisible, even though the defects are reflectionless to Bloch waves \cite{ref1,ref2,ref3,ref4}. This is because, even in case of reflectionless defects \cite{ref1,ref2}, perturbation of the periodic potential introduces some phase delay of Bloch waves, which is not spectrally flat and thus can be detected from e.g. time-of-flight measurements. So far, invisible defects in the bulk of a crystal have been theoretically predicted solely in {\it non-Hermitian} lattices \cite{Longhi1,Longhi2,Longhi3}. Such results would suggest that invisible Tamm states can not be found in an Hermitian lattice. In this work we show that, contrary to such a belief, invisible surface defects sustaining Tamm states can be found in a class of engineered and {\it Hermitian} tight-binding lattices (see section 3). Tamm states sustained by invisible surface defects show a lower-than-exponential localization and have an energy embedded in the allowed energy band of the lattice, i.e. they are bound states in the continuum of von Neumann - Wigner type \cite{Wigner}. While surface Tamm states of BIC type have been theoretically predicted and experimentally observed in recent works \cite{BIC1,BIC2},  such previous surface states were not invisible.     

\begin{figure}
\resizebox{0.48\textwidth}{!}{
\includegraphics{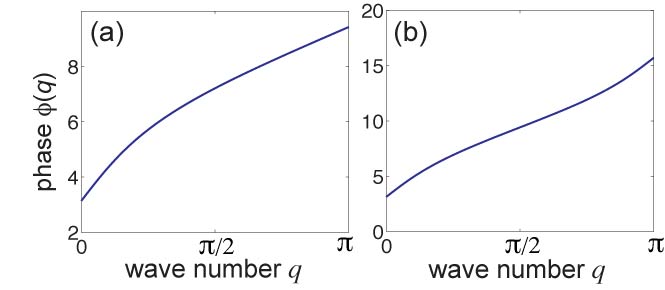}}
\caption{Behavior of the phase $\phi(q)$ of the spectral reflection coefficient $r(q)$ of Bloch waves in (a) the semi-infinite Goodwin's lattice, and (b) the semi-infinite lattice with inhomogeneous hopping rate. Parameter values are $\rho=V_1/ \kappa=2$ in (a) and $\rho= \kappa_0 / \kappa=2$ in (b).}
\end{figure}

\section{Surface Tamm states in simple tight-binding lattice models}
 
\subsection{The Goodwin's model}
One among the simplest models used to explain the appearance of Tamm states with an energy in the gap is the Goodwin's model  \cite{Goodwin1,Goodwin2,Goodwin3,Goodwin4}, which is  discussed in many textbooks as an introductory model of surface electronic states (see, for instance, \cite{book1,book2}). In this model, the electronic states are calculated in a semi-infinite linear chain of atoms with only nearest-neighbor hopping under the tight-binding approximation, the effect of lattice truncation being considered in the change of the site energy of the edge site due to local deformation of the confining potential. The tight-binding Hamiltonian of the Gooddwin's model reads [see Fig.2(a)]
\begin{equation}
\mathcal{H}=\sum_{n=1}^{\infty} \kappa \left( | n \rangle \langle n | +| |n+1 \rangle \langle n|  \right)+\sum_{n=0}^{\infty} V_n | n \rangle
\end{equation}
where $\kappa$ is the hopping rate between adjacent lattice sites, $|n \rangle$ is the Wannier state at site $n$,  and $V_n$ is the potential at site $n$, with $V_n=0$ for $n \neq 1$. The potential $V_1 \neq 0$ at the boundary site $n=1$ differs from the potential of the other sites (here assumed zero as a reference level) owing to lattice truncation. The energy spectrum and corresponding eigenstates of $\mathcal{H}$ are obtained from the lattice equations
\begin{eqnarray}
Ec_n & = & \kappa(c_{n+1}+c_{n-1}) \; \; n \geq 2 \nonumber \\
Ec_1 & = & V_1 c_1+\kappa c_2 
\end{eqnarray}
where $c_n$ is the amplitude probability to find the electron at the Wannier site $|n \rangle$ ($n=1,2,3,....$). For $|V_1|> \kappa$ a bound state, exponentially localized near the surface,  is found with an energy $E=\kappa( V_1/ \kappa + \kappa/ V_1)$, which lies outside (either above or below) the tight-binding lattice band $ -2 \kappa \leq E \leq 2 \kappa$. This is the surface Tamm state, which is therefore a bound state outside the continuum (BOC). The scattered (Bloch-like) states of the lattice Hamiltonian (1) are of the form
\begin{eqnarray}
c_n= \left\{
\begin{array}{lr}
\exp(iqn)+r(q) \exp(-iqn) & \;\; n \geq 2 \\
A & n=1
\end{array}
\right.
\end{eqnarray}
where $ 0 < q < \pi$ is the Bloch wave number (quasi momentum), $r(q)$ is the reflection coefficient and $A$ is the occupation amplitude of the edge site $n=1$. The scattered state (3) has an energy $E=2 \kappa \cos q$ and can be viewed as the superposition of a backward propagating Bloch wave, with quasi momentum $q$ and unit amplitude incident onto the surface,  and a forward propagating (reflected) Bloch wave with amplitude $r(q)$. The reflection coefficient $r(q)$  can be readily calculated and reads
\begin{equation}
r(q)=-\frac{1- \rho \exp(iq)}{1-\rho \exp(-i q)}
\end{equation}
where we have set $\rho \equiv V_1 / \kappa$. Note that for $\rho=0$, i.e. for a defect-free lattice, one has $r(q)=-1$ and $c_n= \sin (nq)$. Conversely,  for $\rho \neq 1$ one has $|r(q)|=1$, corresponding to the fact that the incident wave is fully reflected at the lattice edge, however the phase $\phi(q)$ of $r(q)$ is not flat; see Fig.1(a). This implies that an observer, far from the lattice surface, can detect the presence of the defect by e.g. a time-of-flight measurement. For instance, if an observer placed at the lattice site $n=N \gg 1$ send at initial time $t=0$ a Bloch wave packet with carrier momentum $q=q_0$, the reflected wave packet is observed after the time (so-called phase time \cite{Longhi1})
\begin{equation}
\tau=\tau_0+\frac{1}{v_g}\left( \frac{\partial \phi}{\partial q}\right)_{q_0}
\end{equation}
where $\tau_0=2N/|v_g|$ is the flight-of-time in the absence of the defect at the surface, $v_g=-2 \kappa \sin q_0<0$ is the group velocity of the Bloch wave packet with carrier quasi momentum $q_0$, and $\phi(q)$ is the phase of the reflection coefficient $r(q)$. Equation (5) shows that the surface defect is invisible to an outsider observer if and only if the phase $\phi(q)$ is flat. Since the surface Tamm state in the  Goodwin's model exists provided that $|\rho|>1$, from Eq.(5) it follows that the Tamm state is not invisible to an outsider observer. As an example, in Fig.2(a) we show the numerically-computed propagation of an initial Gaussian-shaped Bloch wave packet, which is reflected from the lattice edge.  The results are obtained by numerically solving the time-dependent Schr\"{o}dinger equation $ i \partial_{t} | \psi (t)\rangle= \mathcal{H} | \psi (t) \rangle$ for the single-particle state $| \psi(t)=\sum_{n=1}^{\infty} c_n(t) | n \rangle$ with the initial condition
\begin{equation}
c_n(0) \propto \exp \left[ -(n-N)^2/w^2+iq_0 n \right]
\end{equation}
where $w$ is the wave packet size, $N$ measures the distance from the edge $n=1$ of the lattice (with $N \gg w $), and $q_0$ is the carrier wave number. The figure compares the case of a defect-free lattice [$V_1=0$, i.e. $\rho=0$, open dots in the right panel of Fig.2(a)] with that of a lattice sustaining a Tamm state  ($\rho=2$, stars).  A spatial shift $\Delta n$ between the two wave packets is clearly observed, which is related to the phase gradient via the relation
\begin{equation}
\Delta n= \left( \frac{\partial \phi}{\partial q} \right)_{q_0}.
\end{equation}  
\begin{figure}
\resizebox{0.45\textwidth}{!}{
\includegraphics{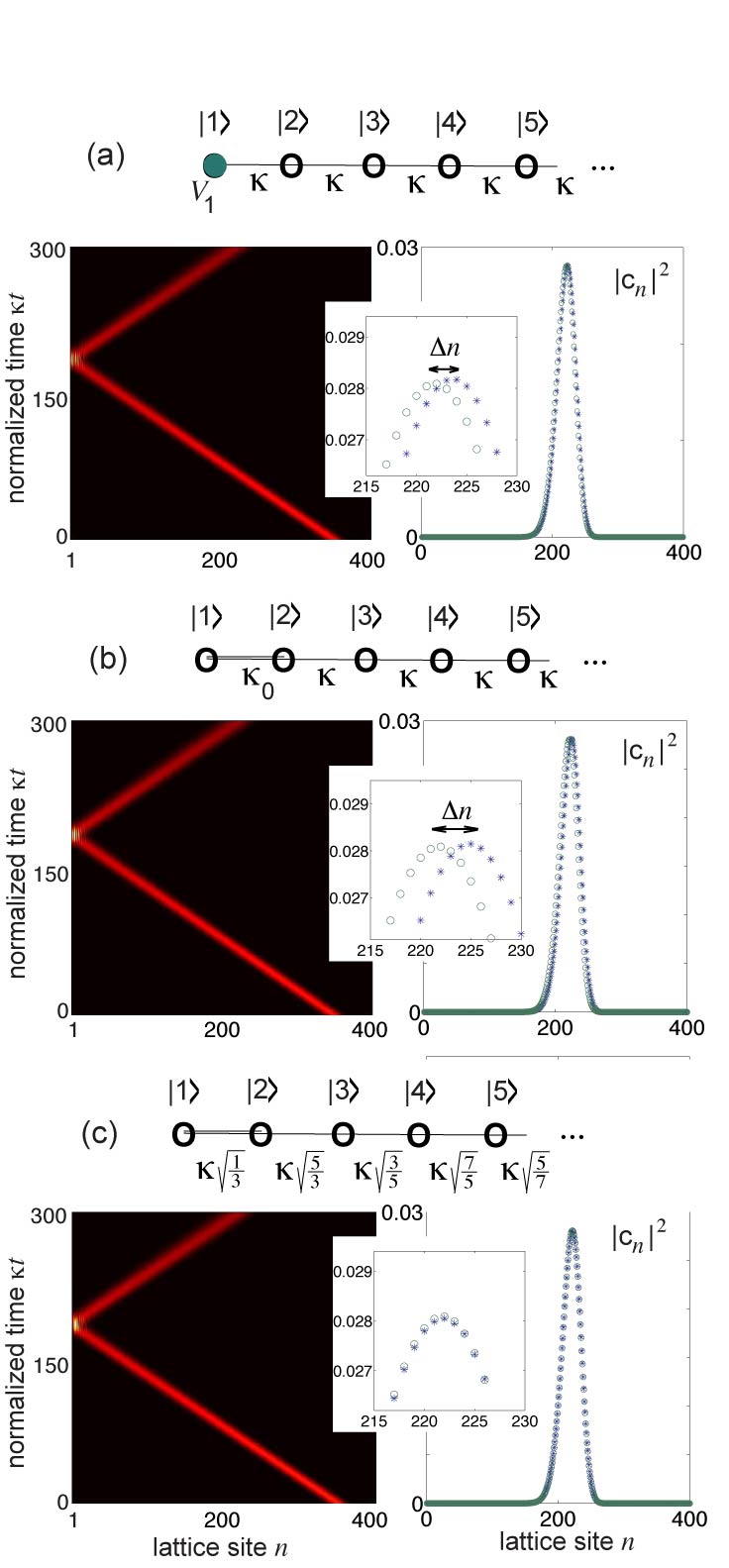}}
\caption{Reflection of a Gaussian wave packet in three different tight-binding semi-infinite lattice models sustaining surface Tamm states: (a) the Goodwin's semi-infinite lattice, (b) semi-infinite lattice with inhomogeneous hopping rate, and (c) semi-infinite lattice with invisible Tamm state. The left panels show the evolution of the site occupation probabilities $|c_n(t)|^2$ in a pseudo-color map. The right panels show the distribution of $|c_n(t)|^2$ at normalized time $\kappa t=300$ (stars); for comparison, the wave packet distribution that one would observe from reflection in the homogeneous (defect-free) semi-lattice is shown by open circles. The insets show an enlargement of the wave packet distribution near the peak. A clear advancement $\Delta n$ is observed in (a) and (b), but not in (c). The initial wave packet distribution is Gaussian-shaped and defined by Eq.(6) with $w=15$, $N=350$ and $q_0=0.4 \pi$.}
\end{figure}

\subsection{Lattice with inhomogeneous hopping rate}
The result found for the Goodwin's model  is a rather general one, i.e. quite generally it turns out that a defect introduced in the semi-infinite lattice near the surface yields a non-flat phase $\phi(q)$ of the reflection coefficient, thus preventing the possibility to realize invisible Tamm states. Here we would like to discuss another simple tight-binding lattice model, where surface Tamm states are sustained by the introduction of inhomogeneous hopping rate (rather than site energy) at the edge of the chain. The Hamiltonian 
of this lattice is given by [see Fig.2(b)]
\begin{equation}
\mathcal{H}=\kappa \sum_{n=2}^{\infty} \left (| n \rangle \langle n+1|+ | n+1 \rangle \langle n | \right)+\kappa_0  \left (| 1 \rangle \langle 2|+ | 2 \rangle \langle 1 | \right)
\end{equation}
and the corresponding tight-binding equations read
\begin{eqnarray}
Ec_n & = & \kappa ( c_{n+1}+ c_{n-1}) \; \; n \geq 3 \nonumber \\
Ec_2 & = & \kappa c_3 + \kappa_0 c_1 \\
Ec_1 & = & \kappa_0 c_2 
\end{eqnarray}
where $\kappa_0$ in the modified hopping rate between sites $n=1$ and $n=2$. 
After setting $\rho= \kappa_0 / \kappa$, it can be readily shown that the lattice sustains two surface Tamm states with energies outside the tight-binding lattice band, i.e. of BOC type, provided that $\rho> \sqrt{2}$. Moreover, the reflection coefficient of Bloch waves is given by
\begin{equation}
r(q)=-\frac{1+(1- \rho^2) \exp(2iq)}{1+(1-\rho^2) \exp(-2iq)}
\end{equation}
A typical behavior of the phase $\phi(q)$ of $r(q)$ is shown in Fig.1(b). Equation (11) shows that, like in the Goodwin's model of surface Tamm states, the phase $\phi(q)$ is not flat and this implies that the surface defect is not invisible. The reflection of a Gaussian wave packet from the surface of this lattice is shown in Fig.2(b), which clearly shows a deviation (delay) from the reflection of the same wave packet by a defect-free surface.    
\begin{figure}
\resizebox{0.48\textwidth}{!}{
\includegraphics{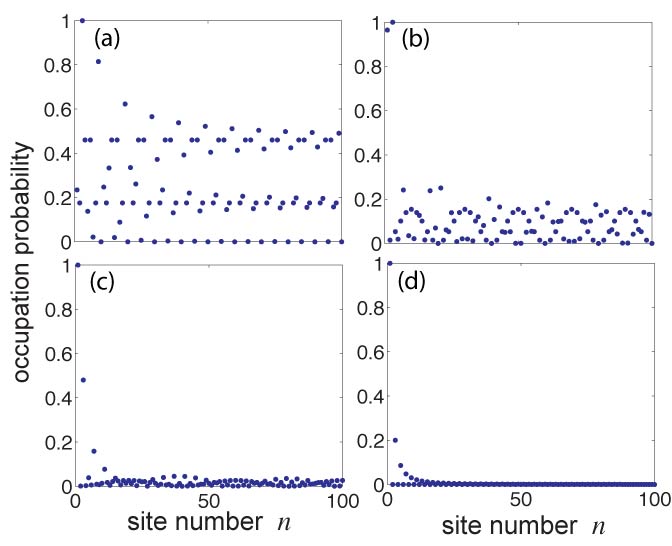}}
\caption{Behavior of the eigenfunction $|c_n|^2$ of $\mathcal{H}$, normalized to its maximum value, as given by Eq.(15), for a few values of $q= \delta \pi/2$: (a) $\delta=0.8$, (b) $\delta=0.9$, (c) $\delta=0.95$, and (d) $\delta=1$. As $\delta$ approaches 1, the scattered state becomes a surface resonance and a truly surface bound state at $\delta=1$.}
\end{figure}

\section{Invisible surface defects}
A main open question is whether one can synthesize a tight-binding lattice with a defect at the surface, which in addition to sustain surface states turns out to be invisible for an outsider observer. Such a question has been recently addressed in Refs.\cite{Longhi1,Longhi2,Longhi4} for defects in the {\it bulk} of a  crystal, where it was shown that invisible defects can be realized provided that {\it non-Hermitian} terms are allowed in the lattice Hamiltonian. The main and distinctive result of the present work is that invisible surface defects, sustaining Tamm states, can exist in a certain class of {\it Hermitian} lattices. 
\subsection{Lattice synthesis}
The strategy is to start from the Hamiltonian of the defect-free semi-infinite tight-binding lattice [Eq.(1) with $V_n=0$], and to synthesize, by application of a double discrete Darboux transformation, an isospectral lattice Hamiltonian $\mathcal{H}$ that sustains a surface BIC state with energy $E=0$ at the center of the lattice band. The technique of double discrete Darboux transformation for tight-binding lattices is rather cumbersome and it is described in details in Appendix A of Ref.\cite{Longhi4}. Its application to the Hamiltonian of the defect-free semi-infinite lattice follows the same lines detailed in Ref.\cite{Longhi4} for the infinitely-extended lattice. With this procedure,
the following isospectral {\it Hermitian} partner Hamiltonian is obtained \cite{note}
\begin{equation}
\mathcal{H}= \sum_{n=1}^{\infty} \kappa_{n+1} \left( | n \rangle \langle n+1|+ | n+1 \rangle \langle n| \right) 
\end{equation}
where  the hopping rates  $\kappa_n$ ($n=2,3,4, ...$) between site $(n-1)$ and site $n$ are given by
\begin{equation}
\kappa_n= \left \{
\begin{array}{ll}
\kappa \sqrt{\frac{n-1}{n+1}} & \;\;\; n \; \; {\rm even} \\
\kappa \sqrt{\frac{n+2}{n}} & \;\;\; n \; \; {\rm odd}.
\end{array}
\right.
\end{equation}
i.e. $\kappa_n / \kappa =\sqrt{1/3}$, $\sqrt{5/3}$, $\sqrt{3/5}$, $\sqrt{7/5}$, $\sqrt{5/7}$, $\sqrt{9/7}$, ... for $n=2,3,4,5,6,7,...$ [see Fig.2(c)]. 
Note that  $\kappa_n \rightarrow \kappa$ as $ n \rightarrow \infty$, i.e. the lattice is asymptotically homogeneous. The stationary states of the Hamiltonian (12) 
satisfy the coupled equations
\begin{eqnarray}
Ec_n & = & \kappa_{n+1} c_{n+1}+ \kappa_{n} c_{n-1} \; \; n \geq 2 \nonumber \\
Ec_1 & = & \kappa_2 c_2. 
\end{eqnarray}
By direct calculations, one can readily show that the scattering state solutions to Eq.(14) with quasi-momentum $q$ and energy $E=2 \kappa \cos q$ are given by
\begin{equation}
c_n=\left\{
\begin{array}{ll}
\sin (qn) & \;\;\; n \; \; {\rm even} \\
\frac{n+1}{\sqrt{(n+1)^2-1}} \sin (qn) -\frac{ {\rm tan} q}{\sqrt{(n+1)^2-1}} & \;\;\; n \;\; \rm {odd}
\end{array}
\right.
\end{equation}
with $q \neq \pi/2$. Note that $c_n \rightarrow \sin (qn)$ as $ n \rightarrow \infty$, so that $r(q)=-1$ for the reflection coefficient. This means that the lattice with hopping rates defined by Eq.(13)  has the {\it same} reflection coefficient of the defect-free semi-infinite lattice, and thus an outsider observer far from the lattice edge can not detect the presence of the defects. For $q \rightarrow \pi/2$, the scattered state (15) becomes a bound state, namely
\begin{equation}
c_n \propto \left\{
\begin{array}{ll}
0 & \;\;\; n \; \; {\rm even} \\
\frac{1}{\sqrt{(n+1)^2-1}} & \;\;\; n \;\; \rm {odd}
\end{array}
\right.
\end{equation}
which is a normalizable state  with a lower than exponential (algebraic) localization; see Fig.3. The energy of this surface Tamm state is $E=0$, i.e. it is a BIC state \cite{note}. This kind of surface Tamm state is analogous to the BIC surface mode recently predicted and observed in Ref.\cite{BIC2} for an Hermitian lattice defined by Eq.(12), but with a different sequence of hopping rates. There is, however, a deep difference between the model of Ref.\cite{BIC2} and the lattice with hopping rates defined by Eq.(13): while in the former case in addition to a BIC state the lattice sustains several other BOC surface states, in the lattice defined by Eqs.(12) and (13) there is only one bound (surface) state. The presence of additional BOC surfaces states in the lattice of Ref.\cite{BIC2} introduces a non-flat phase $\phi(q)$ of the reflection coefficient, like in the models discussed in the previous section, making the surface defect visible to an outsider observer. The invisibility of the surface Tamm state in the ideal lattice, defined by Eq.(13), has been checked by direct numerical simulations, see Fig.2(c). The figure clearly shows that the reflection of an initial Gaussian wave packet, launched at initial time from a position $N$ far from the edge of the surface, turns out to be almost indistinguishable from the reflection of the same wave packet from a defect-free semi-infinite lattice.  
 
\begin{figure*}
\resizebox{0.8\textwidth}{!}{
\includegraphics{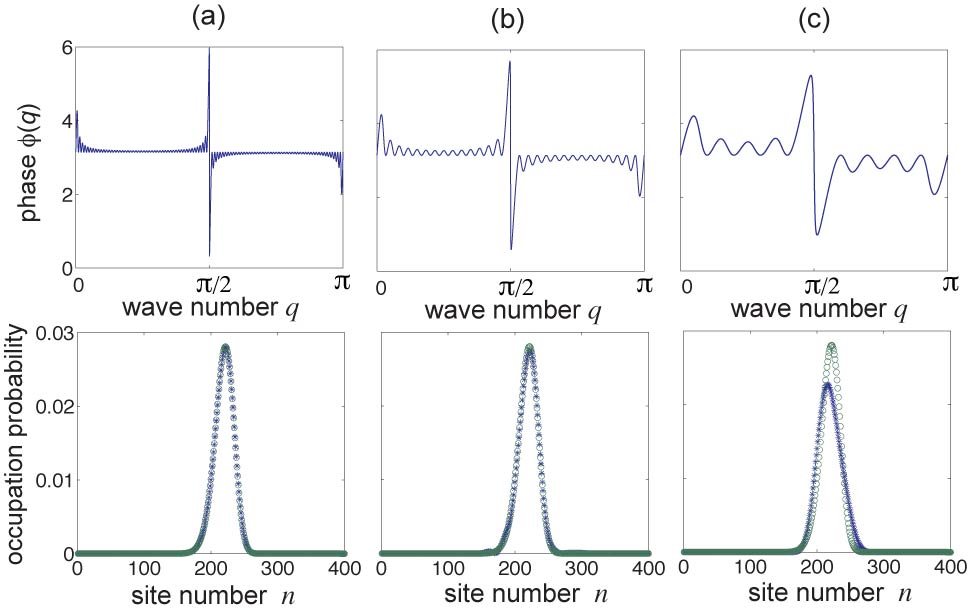}}
\caption{Reflection from a non-ideal lattice with hopping rates given by Eq.(13) for $n \leq N_g$ and $\kappa_n=\kappa$ for $n>N_g$. The upper panels show the numerically-computed phase $\phi(q)$ of the reflection coefficient, whereas the lower panels show the occupation probabilities $|c_n(t)|^2$ at normalized time $\kappa t=300$ for an initial Gaussian wave packet as in Fig.2.  Stars refer to the non-ideal lattice, open circles to the homogeneous (defect-free) lattice. In (a) $N_g=100$, in (b) $N_g=30$ and in (c) $N_g=10$.}
\end{figure*}
\begin{figure*}
\resizebox{0.8\textwidth}{!}{
\includegraphics{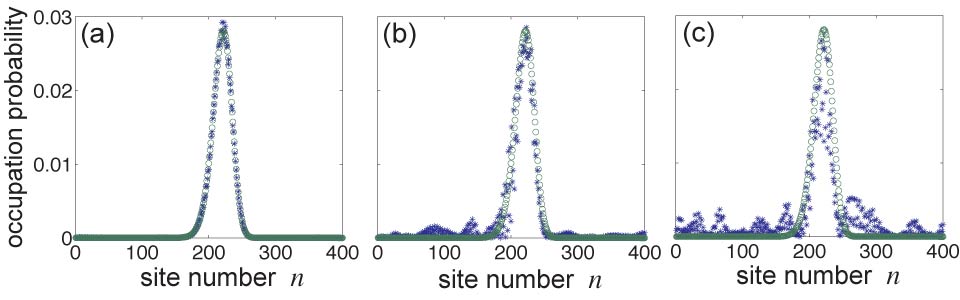}}
\caption{Reflection of a Gaussian wave packet from a disordered semi-lattice for increasing strength $\delta$ of disorder: (a) $\delta=0.01$, (b) $\delta=0.05$ and (c) $\delta=0.1$. The other parameter values are as in Fig.2(c).}
\end{figure*}

\subsection{Effects of lattice imperfections and disorder}
The invisibility of the surface state for the lattice model described in the previous subsection requires a semi-infinite lattice with the precise 'magic' sequence of hopping rates defined by Eq.(13). Lattice imperfections or disorder are expected to destroy the BIC (surface) state, which decays into a {\it resonance} surface state  [i.e. a scattered mode similar to the state plotted in Fig.3(c)]. The fragility of BIC states is a rather general feature, not specific to our model (see, for instance, \cite{cazz2,sci}), however it should be considered for a possible observation of invisible Tamm states. In addition to destroy the BIC state, lattice imperfections are also expected to break the invisible properties of the surface defect. To check the sensitivity of the invisible Tamm state to deviations from the ideal case, in a first set of numerical simulations we studied the reflection properties of a semi-infinite lattice with hopping rates defined by Eq.(13) for $n \leq N_g$, and $\kappa_n=\kappa$ for $n>Ng$. This lattice basically deviates from the ideal one because it is homogeneous for $n>N_g$. This lattice does not sustain anymore a truly bound (surface) Tamm state, i.e. the energy $E=0$ now belongs to the continuous spectrum of $\mathcal{H}$ but corresponds to a resonance state. The reflection coefficient $r(q)$ for this non-ideal lattice can be numerically computed by a standard transfer matrix method. Figure 4 shows, as an example, the behavior of the phase $\phi(q)$ of the reflection coefficient for a few decreasing values of $N_g$. Note that, as $N_g$ decreases, the ideal flat behavior of the phase is lost, with the appearance of oscillations around the mean value $\phi(q)=\pi$ of the invisible case. The oscillations are especially strong is a narrow interval near $q=\pi/2$, i.e. near the resonance states, which broadens as $N_g$ is decreased. In the figure, reflection of a Gaussian wave packet from the lattice surface is also shown and compared to the case of a defect-free surface. Clearly, degradation from the ideal behavior  is observed as $N_g$ is decreased below $ \sim 30$. In a second set of numerical simulations we considered a disordered lattice with hopping rates  given by $\kappa^{'}_{n}=\kappa_n+\kappa  \sigma_n \delta$, where $\kappa_n$ is the ideal distribution given by Eq.(13), $\sigma_n$ is a random number uniformly distributed in the range $(-0.5,0.5)$, and $\delta$ measures the strength of disorder.  Typical results showing the propagation of a Gaussian wave packet for increasing values of the disorder strength $\delta$ are shown in Fig.5. The numerical results indicate that a nearly undistorted reflection, as compared to the defect-free lattice, requires a disorder strength smaller than $\sim 0.01$.\par
As a final comment, it is worth mentioning that, besides structural instability of the invisible defects discussed above, effects of the nonlinearity, e.g. anharmonic distortion of the propagating lattice waves, could also  break the invisibility of the surface defects, even though the nonlinear surface state is expected to survive in the form of so-called embedded soliton (see, for instance, \cite{BIC1,Malomed3} and references therein). In particular, addition of a small amount of focusing or defocusing Kerr nonlinearity is expected to shift the energy of the embedded state inside the lattice band \cite{BIC1}. Nevertheless, anharmonicity and nonlinear effects are fully negligible in case of e.g. surface defects in optical waveguide arrays probed by low-power optical beams \cite{BIC2,ref4}, where breakdown of defect invisibility can arise solely from structural deviations of the static lattice parameters from the ideal ones.

 \section{Conclusions and discussion}
   
Surface Tamm states at the edge of a one-dimensional lattice are threshold states that arise in presence of a sufficiently strong surface defect and generally correspond to bound states with exponential localization and energy in a gap of the crystal. Surface defects influence the reflection of propagative (Bloch) waves, by the addition of a phase which is not flat but depends on the quasi-momentum of the incident wave. This makes the defect not invisible to an observer far from the lattice surface. In this work we have shown that in a certain class of tight-binding lattices surface defects sustaining localized Tamm states can appear invisible to an observer located far from the surface. Tamm states have in this case an energy embedded in the continuous spectrum of the crystal, i.e. they are BIC states of  von Neumann -Wigner kind, and show a lower than exponential localization. Such a result is a non-trivial one and rather surprising, because for infinitely-extended Hermitian lattices defects sustaining localized (defect) modes in the bulk can be at most transparent, but not invisible \cite{Longhi1}. We also discussed the role of lattice imperfections and disorder on the reflection of Bloch wave packets, and shown than the decay of the BIC Tamm state into a resonance surface state induced by lattice imperfections is associated to a breakdown of defect invisibility. It is envisaged that the kind of invisible surface defects sustaining BIC Tamm states predicted in this work could be experimentally observed in a lattice of evanescently-coupled optical waveguides with tailored hopping rates \cite{BIC2}. Also, the use of photonic lattices to investigate Tamm states could be extended to investigate nonlinear phenomena at the surface, for example the occurrence of {\it nonlinear} BIC modes -so called embedded solitons \cite{Malomed3,Malomed1,Malomed2,Malomed4} - and the invisibility/transparency properties of surface defects in the presence of nonlinearity.

\end{document}